\documentclass[fleqn,10pt]{wlscirep}
\usepackage{bm}
\usepackage{amsmath}
\usepackage{color}
\title{Rebuilding of destroyed spin squeezing in noisy environments}

\author[1]{Peng Xu}
\author[1]{Huanying Sun}
\author[2]{S. Yi}
\author[1,*]{Wenxian Zhang}
\affil[1]{School of Physics and Technology, Wuhan University, Wuhan, Hubei 430072, China}
\affil[2]{CAS Key Laboratory of Theoretical Physics, Institute of Theoretical Physics, Chinese Academy of Sciences, P.O. Box 2735, Beijing 100190, China}

\affil[*]{Corresponding email: wxzhang@whu.edu.cn}

\begin{abstract}
We investigate the process of spin squeezing in a ferromagnetic dipolar spin-1 Bose-Einstein condensate under the driven one-axis twisting scheme, with emphasis on the detrimental effect of noisy environments (stray magnetic fields) which completely destroy the spin squeezing. By applying concatenated dynamical decoupling pulse sequences with a moderate bias magnetic field to suppress the effect of the noisy environments, we faithfully reconstruct the spin squeezing process under realistic experimental conditions. Our noise-resistant method is ready to be employed to generate the spin squeezed state in a dipolar spin-1 Bose-Einstein condensate and paves a feasible way to the Heisenberg-limit quantum metrology.
\end{abstract}

\begin{document}

\flushbottom
\maketitle
%
%
\thispagestyle{empty}


\section*{Introduction}

Spin squeezing and quantum entanglement play a critical role in quantum computing~\cite{Horodecki2009Quantum,Bigelow2001Quantum,Haffner2008Quantum}, quantum information science~\cite{Sorensen2001Many,Sorensen2001Entanglement,Korbicz2005Spin,Pezze2009Entanglement}, and quantum metrology beyond the standard quantum limit~\cite{WINELAND1992SPIN,WINELAND1994SQUEEZED,Giovannetti2004Quantum,Leibfried2004Toward,Roos2006Designer,
Polzik2008Quantum,Gross2010nonlinear,Riedel2010atom,Luo2017Deterministic}. During the past two decades, much effort has been devoted to realize spin squeezed and quantum entangled states in various physical systems, such as trapped ions~\cite{Blatt2008Entangled,Svandal2002Collapse}, and photonic systems~\cite{Pan2012Multiphoton}. However, these delicate quantum states are difficult to prepare and easy to lose their quantum coherence, due to the weak coupling to the environments.

A Bose-Einstein condensate (BEC) in ultracold atomic Bose gases shows another great potential in spin squeezing by utilizing the pseudospin or spin degrees of freedom, due to the fine tunability and controllability of the parameters and properties of the system. Recently, experimental realizations of spin squeezing beyond the standard quantum limit, through the one-axis twisting (OAT) which has a theoretical scaling of $1/N^{2/3}$ for $N$ particles~\cite{KITAGAWA1993SQUEEZED}, were reported in two-component BECs~\cite{Gross2010nonlinear,Riedel2010atom}. However, the complete spin squeezed states in the Heisenberg limit, in particular for a large number of spins by two-axis twisting (TAT) method which has a scaling of $1/N$ for $N$ particles~\cite{KITAGAWA1993SQUEEZED}, have not been realized experimentally, because of the challenge to find a suitable system where the TAT method is directly applied.

In pursuing the perfect spin squeezed states in the Heisenberg limit, many theoretical proposals have been developed, including a wonderful theoretical proposal to dynamically generate the TAT Hamiltonian from the OAT one~\cite{Liu2011Spin,Zhang2014Dynamical,Yong2015Quantum}. We refer this method as driven-OAT. Among many candidates for perfect spin squeezing, a more natural one is a dipolar spinor BEC with real atomic spin, such as $^{87}$Rb spin-1 condensates where the coherence time of these systems is extremely long in principle~\cite{Kajtoch2016Spin}. Although the stray magnetic fields in a practical experiment are believed to severely prevent the realization of the perfect spin squeezed states in such systems~[Private communications with M. S. Chapman.], it has not been discussed how the noisy environments affect the spin squeezing dynamics and how to suppress their negative effects.

In this paper, we first briefly review the driven-OAT spin squeezing dynamics in a $^{87}$Rb ferromagnetic dipolar spin-1 BEC, where the magnetic dipolar interaction between atoms is employed to generate the driven-OAT Hamiltonian. We then investigate systematically the stray magnetic fields' effect on the dynamics of the spin squeezing. Indeed, the stray magnetic fields in the order of microGauss completely destroy the spin squeezing. In order to suppress the effect of the noisy environments, we further propose to use the concatenated dynamical decoupling (CDD) pulse sequences with a bias magnetic field and faithfully rebuild the driven-OAT spin squeezing dynamics. Our results show that the dynamical decoupling method is experimentally practical to significantly suppress the noisy environments and effectively revives the spin squeezed state in a dipolar spin-1 BEC.

\section*{Spin squeezing dynamics in a dipolar spin-1 BEC}
\label{sec:sys}

We consider a trapped dipolar spin-1 BEC whose Hamiltonian is~\cite{Ho1998Spinor,Law1998Quantum,Yi2004Quantum,Yi2006Magnetization,
Huang2012Macroscopic,Ning2012Manipulating,Kajtoch2016Spin},
\begin{eqnarray}
\label{eq:OH}
H=H_0+H_d.
\end{eqnarray}
where $H_d$ represents magnetic dipolar interaction between atoms and $H_0$ is the rest part. They are in the second quantized form
\begin{eqnarray}
H_0&=&\int d\bm{r}\hat{\Psi}_{\alpha}^{\dag}(\bm{r})
    \left[{\left(-\frac{\hbar^{2}\nabla^{2}}{2M}+V_{ext}(\bm{r})\right)\delta_{\alpha\beta}}
    {-g_{F}\mu_{B}\bm{B\cdot F_{\alpha\beta}}}\right]\hat{\Psi}_{\beta}(\bm{r}) \nonumber \\
    && +\frac{c_0}{2}\int d\bm{r}\hat{\Psi}_{\alpha}^{\dag}(\bm{r})\hat{\Psi}_{\beta}^{\dag}(\bm{r})
    \hat{\Psi}_{\alpha}(\bm{r})\hat{\Psi}_{\beta}(\bm{r}) \nonumber \\
    && +\frac{c_2}{2}\int d\bm{r} \hat{\Psi}_{\alpha}^{\dag}(\bm{r})\hat{\Psi}_{\alpha'}^{\dag}(\bm{r})
    \bm{F_{\alpha\beta}}\cdot\bm{F_{\alpha'\beta'}}
    \hat{\Psi}_{\beta}(\bm{r})\hat{\Psi}_{\beta'}(\bm{r}), \nonumber \\
H_d&=&\frac{c_d}{2}
    \int \frac{d\bm{r}d\bm{r'}}{|\bm{r}-\bm{r'}|^3}[\hat{\Psi}_{\alpha}^{\dag}(\bm{r})
    \hat{\Psi}_{\alpha'}^{\dag}(\bm{r'})
    \bm{F_{\alpha\beta}}\cdot\bm{F_{\alpha'\beta'}}
    \hat{\Psi}_{\beta}(\bm{r})\hat{\Psi}_{\beta'}(\bm{r'}) \nonumber \\ && -3\hat{\Psi}_{\alpha}^{\dag}(\bm{r})\hat{\Psi}_{\alpha'}^{\dag}(\bm{r'})
    (\bm{F_{\alpha\beta}\cdot e})(\bm{F_{\alpha'\beta'}\cdot e})
    \hat{\Psi}_{\beta}(\bm{r})\hat{\Psi}_{\beta'}(\bm{r'})]. \nonumber
\end{eqnarray}
where $M$ is the mass of the atom, $\hat \Psi_{\alpha}$ the field annihilation operator for spin component $\alpha=-1,0,+1$, $\bm{e=(r-r')/|r-r'|}$ the unit vector. The trapping potential is $V_{ext}({\bm r})=(M/2)(\omega_x^2 x^2 +\omega_y^2 y^2 +\omega_z^2 z^2)$ with $\omega_{x,y,z}$ being the trap angular frequencies.  ${\bm F}=(F_x, F_y, F_z)$ with $F_{x,y,z}$ being spin-1 matrices. The collisional interaction parameters are $c_0={4\pi\hbar^2(a_0+2a_2)}/{3M}$ and $c_2={4\pi\hbar^2(a_2-a_0)}/{3M}$ with $a_{0(2)}$ being the $s$-wave scattering length of two spin-1 atoms in the combined symmetric channel of total spin 0(2). The dipolar interaction parameter is $c_d={\mu_0 g_{F}^2 \mu_{B}^2}/{4\pi}$ with $\mu_0$ being the vacuum magnetic permeability, $g_F$ the Land$\acute{\rm{e}}$ $g$-factor and $\mu_B$ the Bohr magneton. The repeated indices are summed.

We consider a $^{87}$Rb ferromagnetic spin-1 condensate, where $c_2<0$, $c_0 \gg |c_2|$ and $c_d \leq 0.1 |c_2|$, and the single mode approximation is valid in a small or medium magnetic field~\cite{Yi2002Single,Zhang2003Mean,Zhang2015Coherent}. Under this single mode approximation~\cite{Law1998Quantum,Yi2002Single}, $\hat{\Psi}_{\alpha}(\bm{r})\simeq\phi(\bm{r})\hat{a}_{\alpha}$ with $\phi(\bm{r})$ being a spin-independent spatial mode function and $\hat {a}_{\alpha}$ the annihilation operator of spin component $\alpha$. In an axially symmetric trap, the Hamiltonian in Eq.(\ref{eq:OH}) is remarkably simplified as~\cite{Yi2006Magnetization}
\begin{eqnarray}
\label{eq:H}
H=(c_2'-c_d'){\bm{J}}^{2}+3c_d'(J_z^{2}+\hat{n}_0)-g_F\mu_B\bm{B} \cdot {\bm{J}}.
\end{eqnarray}
where $\hat{n}_0=\hat{a}_0^{\dag}\hat{a}_0$ and $\bm{J}=\sum_{\alpha\beta}\hat{a}_{\alpha}^{\dag}\bm{F}_{\alpha\beta}\hat{a}_{\beta}$ is the total spin operator of the dipolar spin-1 condensate. The rescaled interaction strengths are $c_2'=({c_2/2})\int d\bm{r}|\phi(\bm{r})|^4$ and $c_d'=({c_d}/4)\int d\bm{r}d\bm{r}'|\phi(\bm{r})\phi(\bm{r}')|^2
({1-3\cos^{2}\theta_{\bm{e}}})/|\bm{r}-\bm{r'}|^3$
where $\theta_{\bm{e}}$ is the polar angle of the vector $\bm{r}-\bm{r'}$.

For a typical spin-1 condensate, the number of atoms $N$ is much larger than 1, $N\gg 1$. Under this condition, the term with $\hat n_0$ is proportional to $N$ while the terms with $J_z^2$ and $\bm{J}^2$ are proportional to $N^2$. Thus, we may safely neglect the term with $\hat n_0$~\cite{Xing2016Heisenberg,Kajtoch2016Spin} (We justify this approximation in the Supplemental Materials). Consequently, one finds that $[{\bm J}^2,H]=0$ which means the total angular momentum is a constant. Apart from the constant, the Hamiltonian in zero magnetic field becomes
\begin{eqnarray}
\label{eq:HOAT}
H_{OAT}=\chi J_z^{2}.
\end{eqnarray}
where $\chi=3c_d'$ and hereafter we set $\hbar = 1$. Such a Hamiltonian is nothing but the famous OAT Hamiltonian which is used to generate spin squeezing.

\begin{figure}
\begin{center}
\includegraphics[width=6.5in]{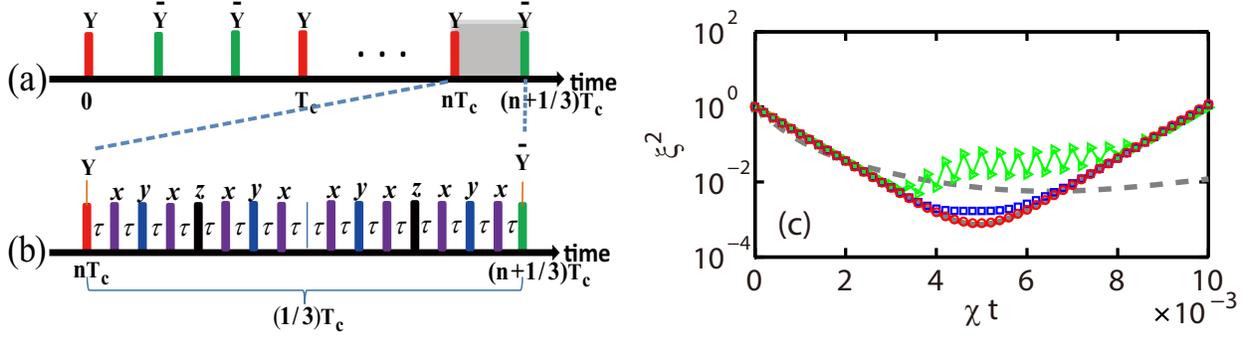}
\end{center}
\caption{\label{fig:pulse} (a) A pulse sequence used to effectively generate TAT from OAT at zero magnetic field. The red (green) bars with $Y$ ($\bar Y$) represent a $\pi/2$ pulse around $y$ ($-y$) axis. The system evolves freely between pulses. $T_c$ is the cycle period. (b) A CDD pulse sequence used to suppress the stray magnetic fields for driven-OAT. The purple (blue, black) bars with $x$ ($y, z$) represent a $\pi$ pulse around $x$ ($y, z$) axis. (c) Spin squeezing for OAT (grey dashed line), TAT (grey solid line), and driven-OAT with $\chi T_c = 4\times10^{-4}$ (green triangles), $2\times10^{-4}$ (blue squares), and $1\times10^{-4}$ (red circles). As $T_c$ decreases, the driven-OAT approaches to the TAT.}
\end{figure}

Since the optimal squeezing direction is very complex to obtain and the optimal scaling of squeezing parameter is $1/N^{2/3}$ for $N$ particles under OAT, one expects to generate the perfect spin squeezed state with a fixed spin squeezing direction and with the Heisenberg limit by employing a TAT Hamiltonian~\cite{KITAGAWA1993SQUEEZED}
\begin{eqnarray}
\label{eq:HTAT}
H_{TAT} = \frac{\chi}{3} (J_x^{2}-J_y^2).
\end{eqnarray}
This Hamiltonian is also known as double quantum Hamiltonian in nuclear magnetic resonance community~\cite{Zhang2009NMR}. The optimal spin squeezing occurs at $t \approx 3 \ln(8N)/(4N) $. The direction of the optimal spin squeezing is along the angle bisector of the $x$ and $y$ axes~\cite{Liu2011Spin}. The TAT Hamiltonian is theoretically proposed to be effectively generated by periodically driving the OAT Hamiltonian~\cite{Liu2011Spin,Zhang2014Dynamical}
\begin{eqnarray}
\label{eq:HD}
H_D= \chi J_z^{2}+ {\pi\over 2} J_y\sum_{n=0}^\infty \left[\delta(t-nT_c)-\delta(t-(n+1/3)T_c)-\delta(t-(n+2/3)T_c)\right].
\end{eqnarray}
where $n$ belongs to integer and the term with $J_y$ describes the spin rotation around the $\pm y$-axis with an angle $\pi/2$ in a form of hard pulse, as shown in Fig.~\ref{fig:pulse}(a). We refer hereafter this Hamiltonian as driven-OAT.

In this paper, we choose the number of $^{87}$Rb atoms $N=1250$ in the spin-1 condensate. The coupling constant $\chi$ is evaluated for a typical optical trap with frequencies $(\omega_x, \omega_y, \omega_z) = 2\pi(150,150,1200)$ Hz. By searching for the ground state of the fully polarized condensate with imaginary time propagation method based on the three coupled Gross-Pitaevskii equations, the quantity for the interaction strength we numerically find $\chi \approx 6.762\times 10^{-4}$ Hz~[Due to strong repulsive interaction described by the term with $c_0$, the condensate density is much lower than the ideal gases' density. Thus, the $\chi$ is also smaller than the value $\chi\approx 2.324\times 10^{-3}$ Hz, obtained analytically by assuming a Gaussian ansatz in Ref.~\cite{Huang2012Macroscopic}.].

We prepare the condensate in a fully polarized initial spin state $|J_z=N\rangle$ with the total spin $J=N$. We define the spin squeezing parameter among the triple of operator spanning the su(2) subalgebra~\cite{Yukawa2013Classification}. Due to the fact that our initial state $|J_z=N\rangle$ is defined with respect to the Bloch sphere of the $\{J_x,J_y,J_z\}$ operator, one can extend the definition of this parameter from the spin-1/2 ensemble to the spin-1 condensate~\cite{KITAGAWA1993SQUEEZED, Ma2011Quantum, Kajtoch2016Spin}
\begin{eqnarray}
\label{eq:SSP}
\xi^2=\frac{2(\Delta J_{\bot})_{\rm min}^2} {J}.
\end{eqnarray}
Here, $(\Delta J_{\bot})_{\rm min}^2$ represents the smallest variance perpendicular to the mean spin direction.

We numerically simulate the evolution of the dipolar spin-1 condensate from the prepared initial state under the OAT, TAT, and driven-OAT Hamiltonians. The results are presented in Fig.~\ref{fig:pulse}(c), similar to the previous results obtained by Liu {\it et al.}~\cite{Liu2011Spin}. Clearly, the optimal spin squeezing parameter of the TAT is about an order of magnitude smaller than that of the OAT. More importantly, the spin squeezing parameter of the driven-OAT approaches that of the TAT if the cycle period $T_c$ is small enough. This condition is a reasonable results from the Magnus expansion,
\begin{eqnarray}
\left[e^{-i\pi/2 J_y} e^{-iT_c\chi J_z^2/3} e^{i\pi/2 J_y} e^{-iT_c\chi J_z^2/3} e^{i\pi/2 J_y} e^{-iT_c\chi J_z^2/3} e^{-i\pi/2 J_y} \right]^n = e^{-i\,nT_c [H_{TAT}+O(J^4\chi^3 (T_c/3)^2)]}. \nonumber
\end{eqnarray}
if $J^2(\chi T_c)^2 \ll 1$. In Fig.~\ref{fig:pulse}(c) with $\chi T_c = 10^{-4}$, since $J^2(\chi T_c/3)^2 \sim 0.002$, the dynamical process of the driven-OAT coincides with the TAT's results, which is consistent with the analysis based on the Magnus expansion.

\section*{Detrimental effect of noisy environments}
\label{sec:mf}

\begin{figure}
\begin{center}
\includegraphics[width=6.5in]{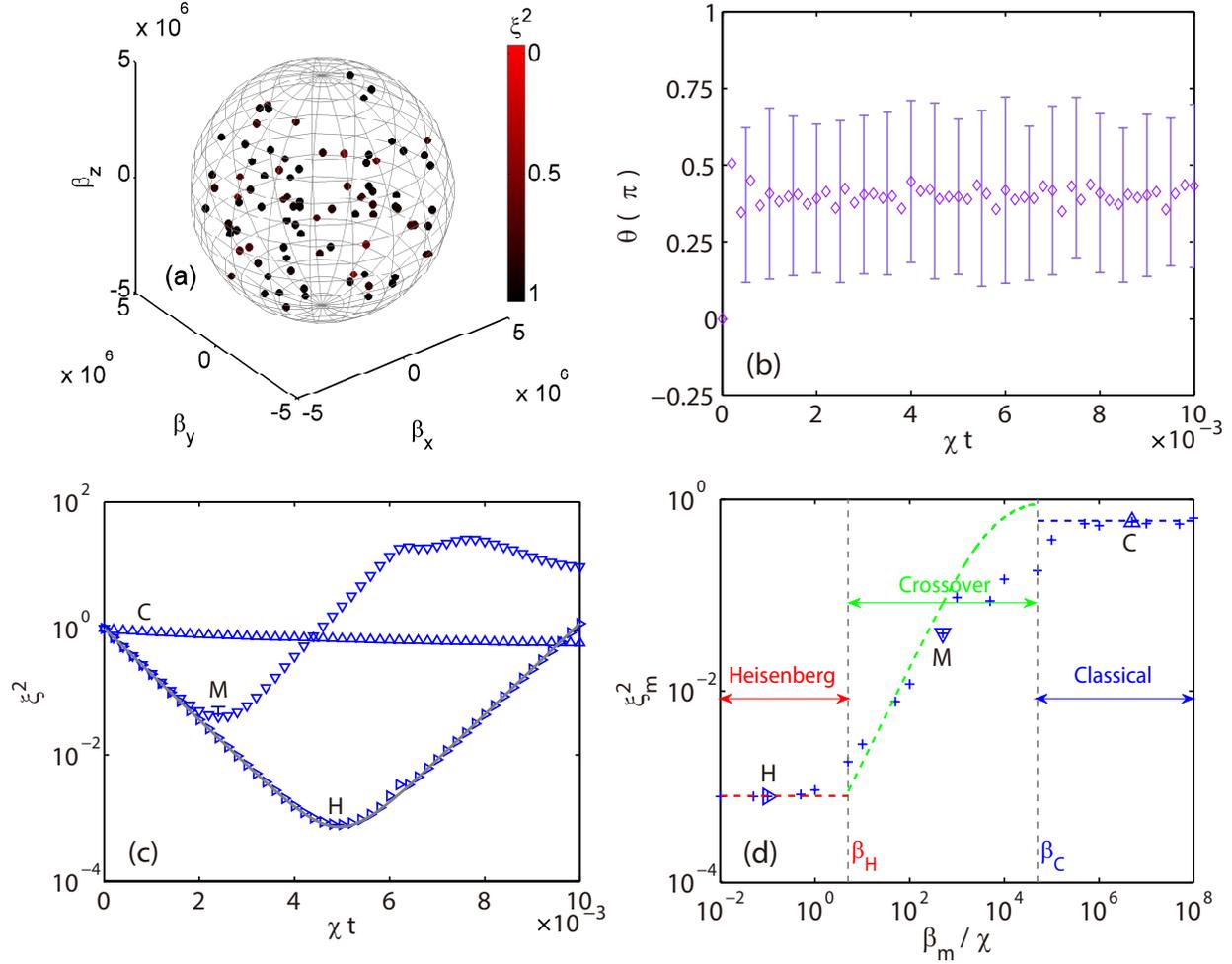}
\end{center}
\caption{\label{fig:mf} (a) Samples of 100 random magnetic fields with the maximum magnitude $\beta_m =5\times 10^6 \chi$ (i.e., $B_m = 4.83$ mG). Color scales the optimal spin squeezing parameter during a driven-OAT process with a given magnetic field. (b) Typical mean spin direction and error bars averaged over 100 samples with $\beta_m = 5\times 10^6 \chi$. The purple diamonds denote the polar angle $\theta$. (c) Averaged evolution of spin squeezing parameter in stray magnetic fields with $\beta_m=5\times 10^6\chi$ (labeled by C, up triangles), $5\times 10^2\chi$ (M, down triangles), and $0.1\chi$ (H, right triangles). The driven-OAT cycle time $T_c=1\times 10^{-4}\chi^{-1}$. The solid grey curve represents the dynamics of TAT. (d) Optimal averaged spin squeezing parameter for different $\beta_m$. The results shown in panel (c) are marked correspondingly with the same type triangles. Three regions are identified as Heisenberg, crossover, and classical, denoting the quantum Heisenberg limit, the crossover, and the classical limit, respectively. The vertical dashed lines with $\beta_H$ and $\beta_C$ mark the boundaries between the three regions. The blue and red horizontal dashed lines are the classical ($\xi_m^2\approx 0.6$) and Heisenberg limit ($\xi_m^2\approx 1/N$), respectively. The green dashed line depicts the analytic results of Eq.~(\ref{eq:xi}) in the crossover region.}
\end{figure}

In a practical spin squeezing experiment of a dipolar spin-1 BEC, there are inevitably noisy environments, e.g., the stray magnetic fields, originated from different sources. These fields are in the order of 1 mG in a common Lab and 0.1 mG in a magnetically shielded room~\cite{Eto2013Spin}. The corresponding Lamor precession rates are about $70$ Hz at 0.1 mG and $700$ Hz at 1 mG for $^{87}$Rb, which are much larger than the dipolar interaction strength for each atom $\sim \chi\, N$. Such a ``strong'' field certainly affects the spin squeezing dynamics and may eventually destroy the spin squeezed state.

In the following, we numerically examine the effect of the stray magnetic fields on the driven-OAT. We can safely neglect the quadratic Zeeman energy on the dynamical squeezing process since it is extremely small. For example, at a field $B_z=4.83$ mG, the quadratic Zeeman shift $q=1.681\times 10^{-3}$ Hz is much smaller than the linear one $p=3.381\times 10^3$ Hz. In addition, the quadratic Zeeman energy is also much smaller than the dipolar interaction $(qN/\chi N^2)\sim 2\times 10^{-3}$. Thus, the Hamiltonian Eq.~(\ref{eq:HD}) in a stray magnetic field $\bm{\beta}\equiv g_F\mu_B {\bm B}$ becomes
\begin{eqnarray}
\label{eq:HDM}
H_{D}=\chi J_z^2 + \bm{\beta} \cdot \bm{J} + {\pi\over 2} J_y\sum_{n=0}^\infty \left[\delta(t-nT_c)-\delta(t-(n+1/3)T_c)-\delta(t-(n+2/3)T_c)\right].
\end{eqnarray}
We assume that the stray magnetic fields are constant for each experimental shot but changes randomly for different shots. The measured properties are actually averaged over many experimental shots. We mimic this quantum process by simulating the spin squeezing over 100 random magnetic fields, which obeys a uniform random distribution in the sphere $|\beta\leq \beta_m|$, as illustrated in Fig.~\ref{fig:mf}(a).

The averaged evolution of the spin squeezing in stray magnetic fields with $\beta_m = 5\times 10^6\chi$ is plotted in Fig.~\ref{fig:mf}(b) for the direction of the mean spin and in Fig.~\ref{fig:mf}(c) for the spin squeezing parameter (up triangles). We find that (i) the mean spin direction is completely random; (ii) no spin squeezing compared to a classical coherent state, i.e., $\xi^2 \sim 1$. Such results are reasonable since the environments are so noisy and destroy significantly the spin squeezing process. No spin squeezing occurs at all in the strong stray magnetic fields.As we decrease $\beta_m$, the spin squeezing emerges gradually and approaches the ideal Heisenberg limit. The moment of the optimal spin squeezing parameter also extends to a longer time.

We investigate systematically the dependence of the optimal spin squeezing parameter on the $\beta_m$ and summarize the results in Fig.~\ref{fig:mf}(d). We observe two plateaus in the weak and strong stray magnetic fields and a crossover in between. The weak field plateau is nothing but the Heisenberg limit from the TAT Hamiltonian, while the strong field plateau shows no-squeezing as the same as the classical coherent state. This plateau-crossover-plateau feature guides us to identify three regions: the Heisenberg region, the classical region, and the crossover. The boundaries are defined approximately as $\beta_H$ where $\xi^2 = 2\xi^2_H$ (i.e., 3 dB above the Heisenberg limit) with $\xi_H^2\approx 1/N$ the Heisenberg plateau value, and similarly $\beta_C$ where $\xi^2 = \xi^2_C/2$ (i.e., -3 dB below the classical limit) with $\xi_C^2\approx 0.6$ the classical plateau value.

The appearance of the classical plateau is easily understood as one follows Law {\it et al.}'s derivation~\cite{Law2001Coherent, Kajtoch2016Spin}. In a strong stray magnetic field, the effective Hamiltonian $e^{i\pi J_y/2} H e^{-i\pi J_y/2}$ in the first interval $T_c/3$ is
\begin{eqnarray}
H_{eff} = \chi J_x^2 - \beta_z J_x + \beta_y J_y + \beta_x J_z.
\end{eqnarray}
The Heisenberg equations of motion for the condensate spin in the x and y directions are~\cite{Law2001Coherent}
\begin{eqnarray}
\dot{J}_x&=& -\beta_x J_y + \beta_y J_z,\nonumber \\
\dot{J}_y&=& -\chi (J_xJ_z+J_zJ_x)+\beta_x J_x + \beta_z J_z. \nonumber
\end{eqnarray}
The solutions are
\begin{eqnarray}
J_x(t)&=&\left[J_x(0)+\frac{\beta_z J}{\beta_x-2\chi J}\right]\cos(\omega t)
-{\beta_x \over \omega} \left[J_y(0)-\frac{\beta_y J}{\beta_x}\right]\sin(\omega t)
-\frac{\beta_z J}{\beta_x-2\chi J}, \nonumber \\
J_y(t)&=&\left[J_y(0)-\frac{\beta_y J}{\beta_x}\right]\cos(\omega t) +
{\omega \over \beta_x} \left[J_x(0)+\frac{\beta_z J}{\beta_x-2\chi J}\right]\sin(\omega t)
+\frac{\beta_y J}{\beta_x}.\nonumber
\end{eqnarray}
where we have adopted the approximation that $J_z\approx J$ in the first interval $T_c/3$ because the initial spin is along $z$ direction and $\omega\equiv \sqrt{(\beta_x)^2-2\beta_x \chi J}$ is defined. The variance of the spin in $x$-$y$ plane becomes
\begin{eqnarray}
\langle(\triangle J_{\phi})^2\rangle
&=&[V_x\cos^2(\omega t)+\frac{(\beta_x)^2}{\omega^2}V_y\sin^2(\omega t)]\cos^2{\phi} \nonumber \\
&+&[V_y\cos^2(\omega t)+\frac{\omega^2}{(\beta_x)^2}V_x\sin^2(\omega t)]\sin^2{\phi}. \nonumber
\end{eqnarray}
where $V_{x,y} = \langle[J_{x,y}(0)]^2\rangle$ and $\phi$ is the azimuthal angle. For the initial coherent state, $V_{x,y}=J/2$ and consequently the spin squeezing parameter becomes
\begin{eqnarray}
\xi^2=\frac{2(\Delta J_{\bot})_{\rm min}^2}{J}\approx 1.
\end{eqnarray}
because $|\beta_x| / \omega \approx 1$ ($|\beta_x| \gg \chi J$). Similar analysis is also applied at longer times. In the cases of extremely large magnetic fields $\beta_m T_c/3 >1$, we split the time interval $T_c/3$ into many smaller intervals $T_c'=T_c/K$ and $\beta_m T_c'/3 \ll 1$. The above analysis is performed similarly for each smaller interval $T_c'/3$. Therefore, in a strong stray magnetic field, the spin squeezing parameter is independent of the field strength $\beta_m$ and is about the same as that for a coherent state. We call the strong field plateau, where the optimal squeezing parameter is approximately $0.6$ [the blue dashed line in Fig.~\ref{fig:mf}(d)], the classical region.

In the weak field limit, there exists another plateau, Heisenberg region,  in Fig.~\ref{fig:mf}(d). The optimal squeezing parameter of the plateau almost equals to $\xi^2=1/N$ (the red dashed line)~\cite{KITAGAWA1993SQUEEZED}, i.e., the stray magnetic fields do not affect the squeezing dynamics. However, we note that the field strength $\beta_H$ is about 5 nano-Gauss, which is impossible to be realized in current $^{87}$Rb spinor condensate experiments.

Between the classical and Heisenberg regions, there exists a crossover region where the optimal squeezing parameter increases monotonically with the increase of the stray magnetic fields $\beta_m$. The feature in the crossover region can be understood by adopting the Magnus expansion, where the effective Hamiltonian in a period of $T_c$ is
\begin{eqnarray}
\label{eq:Heff}
H_{eff}=\frac{1}{3}[\chi(J_x^2-J_y^2)+\beta_x J_x + 3\beta_y J_y + \beta_z J_z)].
\end{eqnarray}
Similar to the classical region, the evolutions of the condensate spin in the $x$ and $y$ directions become
\begin{eqnarray}
J_x(t)&=&A+C, \nonumber \\
J_y(t)&=&B+D. \nonumber
\end{eqnarray}
where
\begin{eqnarray}
A&=& J_x(0) \cosh(\tilde \omega t) - J_y(0) \sqrt{\frac{\omega_1}{\omega_2}} \sinh(\tilde \omega t), \nonumber \\
B&=& J_y(0) \cosh(\tilde \omega t) - J_x(0) \sqrt{\frac{\omega_2}{\omega_1}} \sinh(\tilde \omega t), \nonumber \\
C&=&\frac{\beta_x J}{\omega_2} \cosh(\tilde \omega t) + \frac{\beta_y J}{\tilde \omega} \sinh(\tilde \omega t) -\frac{\beta_x J}{\omega_2}, \nonumber \\
D&=& -\frac{3\beta_y J}{\omega_1} \cosh(\tilde \omega t) - \frac{3\beta_x J}{\tilde \omega} \sinh(\tilde \omega t) +\frac{3\beta_y J}{\omega_1}. \nonumber
\end{eqnarray}
with $\omega_1=2\chi J + \beta_z$, $\omega_2=2\chi J - \beta_z$, and $\tilde \omega = \sqrt{\omega_1 \omega_2}\, /3$. We have assumed $J_z\approx J$  before the optimal squeezing time which is verified numerically. The variance of the spin in the $x$-$y$ plane with an azimuthal angle $\phi$ is
\begin{eqnarray}
\langle(\triangle J_{\phi})^2\rangle &=& V(A)\cos^2(\phi) + V(B)\sin^2(\phi) + Cov(A,B)\sin(2\phi). \nonumber
\end{eqnarray}
where $V(A)$, $V(B)$ are the variances for $A$ and $B$, respectively. $Cov(A,B)$ is the covariance between $A$ and $B$. The constants $C$ and $D$ contribute nothing to the variance $\langle(\triangle J_{\phi})^2\rangle$. Since the optimal spin squeezing direction in TAT at zero field is $\phi = \pi/4$, we calculate the following squeezing parameter
\begin{eqnarray}
\label{eq:xi}
\xi^2&=&\frac{2(\Delta J_{\pi/4})^2}{J}, \nonumber \\
     &\approx& e^{-r} + \frac{t \langle \beta_z^2\rangle}{6\chi J}\, e^{-r} + \frac{\langle \beta_z^2\rangle}{16\chi^2 J^2} \left( 3e^{-r} + e^{r} - 4 \right).
\end{eqnarray}
with $r = 4\chi J t/3$. We have averaged the squeezing parameters over all stray magnetic fields samples so that $\langle \beta_z^2\rangle = \beta_m^2/5$. For a given $\beta_m$, the optimal squeezing parameter $\xi_m^2$ is determined by finding the minimum value at different evolution times. Interestingly, the relation between $\xi_m^2$ and $\beta_m$ is in a simple square law. We plot the result of Eq.~(\ref{eq:xi}) with green dashed line in Fig.~\ref{fig:mf}(d). The analytical results are in good agreement with the numerical results in the crossover region~[The deviation of the analytical results from the numerical ones close to the classical region is due to the breakdown of the first order Magnus expansion Eq.~(\ref{eq:Heff}).].

\section*{Rebuilding spin squeezing with dynamical decoupling}
\label{sec:dd}

\begin{figure}
\begin{center}
\includegraphics[width=6.5in]{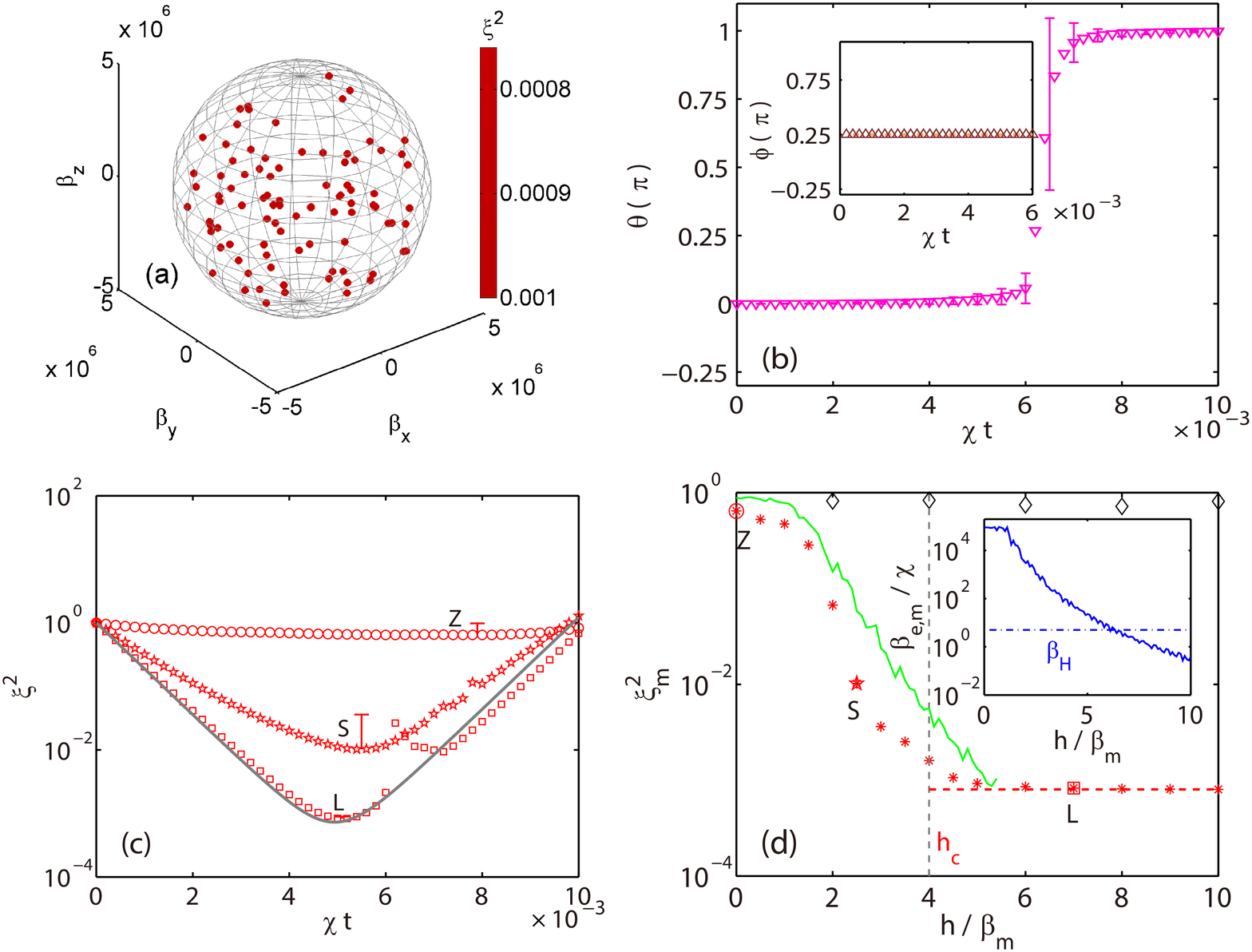}
\end{center}
\caption{\label{fig:dd} Same as Fig.~\ref{fig:mf}(a) and Fig.~\ref{fig:mf}(b), except with the CDD pulse sequence and a large bias $h/\beta_m=7$. Inset shows the azimuthal angle of optimal squeezing direction in the $x$-$y$ plane. (c) Averaged evolutions of spin squeezing parameter under the CDD sequence with the bias $h/\beta_m=0$ (circles), $2.5$ (stars), $7$ (squares). Each curve (with an error bar at optimal squeezing time) is over 100 realizations of stray magnetic fields. The solid grey curve represents the dynamics of TAT. (d) Dependence of the optimal (averaged) spin squeezing parameter $\xi^2_m$ (red asterisks with CDD, grey diamonds without CDD) on the bias $h$. The horizontal red dashed line denotes the Heisenberg limit, $\xi^2_m=1/N$. The green curve represents the analytic result of Eq.~(\ref{eq:xi}). The vertical dashed line with $h_C$ marks a boundary where $\xi^2_m=2/N$. Obviously, the CDD sequence with CDD and a large bias rebuilds the optimal spin squeezing perfectly while a bias alone can not. Inset shows the maximum of the effective stray magnetic fields $\beta_{e,m}$ in a CDD cycle.}
\end{figure}

It is clear from Fig.~\ref{fig:mf} that a small stray magnetic field ($\beta_m \ge \beta_C = 5\times10^4 \chi$, i.e., 50 $\mu$G) completely destroys the spin squeezing dynamics. To generate the squeezed spin state in such noisy environments, one has to suppress significantly the noise. The often employed approach is dynamical decoupling, borrowed from nuclear magnetic resonance community and quantum computing, including periodic dynamical decoupling, symmetrized dynamical decoupling, and CDD~\cite{Viola1999Dynamical, Khodjasteh2005Fault, Zhang2008Long}. With numerical simulations, we find all these dynamical decoupling sequences with hard pulses suppress well the stray magnetic fields' effect with a short enough pulse delay $\tau$. However, in a real spin-1 condensate experiment, the pulse delay is lower bounded by the $\pi$-pulse width, which is in the order of $20$ $\mu$s according to Eto {\it et al.}'s report~\cite{Eto2014Control}, and upper bounded by the the driven-OAT pulse delay $T_c/3$. To satisfy the hard pulse approximation, we must choose $\tau$ as long as possible. Therefore, we set $\tau =\tau_m = T_c/48$, the upper bound in the CDD sequence. For a $^{87}$Rb condensate in a cylindrical trap with the previous given parameters, such a pulse delay $\tau$ is about $490 \;\mu$s (i.e., $T_c=23.52$ ms), much longer than the $\pi$-pulse width. In our calculations, we approximate these $\pi/2$ and $\pi$ pulses as hard pulses (i.e., $\delta$ function with zero width).

At zero bias field, all the dynamical decoupling sequences with $\tau=\tau_m$ unfortunately fail to suppress the stray fields' effect [e.g., Fig.~\ref{fig:dd}(d)]. The failures indicate that we have to apply a bias field $h$. The direction of the bias field must be along the $z$ axis, due to the commutating requirement of the nonlinear squeezing term $\chi J_z^2$. As to the strength of the bias field, one expects a possible small value in order to avoid its quadratic Zeeman effect. Among these DD sequences, the CDD depicted in Fig.~\ref{fig:pulse}(b) with a nonzero bias performs the best.

We investigate systematically the CDD performance in various bias fields at $\tau=\tau_m\approx 2\times 10^{-6}\chi^{-1}$ and present our numerical results in Fig.~\ref{fig:dd}. As shown in Fig.~\ref{fig:dd}(a) and (b), the optimal spin squeezing parameters in each realization of a strong stray magnetic field are almost the same as the zero field case, indicating the significant suppression effect of the CDD sequence with $h/\beta_m = 7$ and $\beta_m=5\times 10^6\chi$. Moreover, the polar $\theta$ of the spin direction and the azimuthal $\phi$ angle of the optimal squeezing direction are nearly the same as the theoretical prediction from the TAT model, i.e., $\theta = 0$ and $\phi = \pi / 4$. We present in Fig.~\ref{fig:dd}(c) the spin squeezing dynamics for three typical biases. In the case of zero bias ($h = 0$, marked with 'Z' in the figure), the spin squeezing parameter changes little with time and the optimal value is about the same as that of a coherent state, indicating the failure of suppressing the stray magnetic fields. In the case of a small bias ($h/\beta_m = 2.5$, marked with 'S'), although we observe quite large spin squeezing, the optimal value is still above the Heisenberg limit. In the case of a large bias ($h/\beta_m = 7$, marked with 'L'), the spin squeezing dynamics is rebuilt almost perfectly, manifesting clearly the significant suppression of the stray magnetic fields by the CDD sequence. To understand more clearly about the bias field effect, we investigate systematically with numerical calculations and plot the dependence of the optimal squeezing parameter on the bias in Fig.~\ref{fig:dd}(d). The three typical cases in Fig.~\ref{fig:dd}(c) are also included and marked with the same letters. It is obvious that the stray magnetic fields effect is suppressed completely and the Heisenberg limit of the spin squeezing parameter is reached, if the bias field is large. While in a small magnetic field, the optimal spin squeezing parameter increases monotonically as the bias decreases. We may define a critical bias $h_c$, which is the bias where $\xi^2_m$ is 2 times as large as (i.e., 3 dB above) the Heisenberg limit $1/N$. For a $^{87}$Rb condensate in a cylindrical trap with the previous given parameters, the critical bias is about $20$ mG, a value to safely neglect its quadratic Zeeman effect.

As shown in Fig.~\ref{fig:dd}, the effect of the noisy environments is significantly suppressed by the CDD sequence in a large bias field. The spin squeezing dynamics of the driven-OAT is rebuilt perfectly. Such a good decoupling enables the condensate spin practically immune to the noisy environments. In the following, we calculate the effective stray magnetic fields to explain our numerical results in a bias magnetic field $h$. The time-modulated Hamiltonian is, in the laboratory reference frame,
\begin{eqnarray}
\label{eq:HDMC}
H_{DD}(t) &=& H_D + hJ_z + H_{CDD}(t).
\end{eqnarray}
where
\begin{eqnarray}
H_{CDD}(t) &=& \pi \sum_{n=0}^\infty \sum_{j=0}^2 \left[J_x \sum_{k=0}^7 \delta(t-(n+j/3)T_c-(2k+1)\tau)
+ J_y \sum_{k=0}^3 \delta(t-(n+j/3)T_c-(4k+2)\tau)
\right. \nonumber \\
&+& \left. J_z \sum_{k=0}^1 \delta(t-(n+j/3)T_c-(8k+4)\tau)
\right]. \nonumber
\end{eqnarray}
denotes the CDD hard pulse sequence. The term $H_D$ in Eq. (8) includes the nonlinear interaction and the stray magnetic fields' effect.

In the toggling frame defined by the CDD pulses~\cite{Haeberlen2012High}, the evolution operator becomes approximately (time ordered) $U_{CDD}=e^{-i 16 \tau \chi J_z^2} \, T: \prod_{j=1}^{16} U_j$, where $U_j = \text{exp}(-i \tau H_j)$ with
\begin{eqnarray}
H_1&=&\beta_x J_x + \beta_y J_y + h'J_z,  \nonumber \\
H_2&=&\beta_x J_x - \beta_y J_y - h'J_z,  \nonumber \\
H_3&=& - \beta_x J_x - \beta_y J_y + h'J_z,  \nonumber \\
H_4&=& - \beta_x J_x + \beta_y J_y - h'J_z,  \nonumber \\
H_5&=& \beta_x J_x - \beta_y J_y - h'J_z,  \nonumber \\
H_6&=& \beta_x J_x + \beta_y J_y + h'J_z,  \nonumber \\
H_7&=& - \beta_x J_x + \beta_y J_y - h'J_z,  \nonumber \\
H_8&=& - \beta_x J_x - \beta_y J_y + h'J_z  \nonumber
\end{eqnarray}
and $H_j = H_{17-j}$ for $j=9, 10, \cdots, 16$. Here, we have assumed $J_z^2$ a constant and introduced $h'=h+\beta_z$~\cite{Law2001Coherent}. To calculate the product of $U_j$'s, the usual average Hamiltonian theory is not applicable because $\beta_{x,y} \tau \lesssim 10$ and $h'\tau \lesssim 100$ beyond the convergence radius~\cite{Takegoshi2015Comparison,Mananga2015Advances,Slichter1990Principles,Haeberlen2012High}. We employ, instead, a continuous rotation method developed by us previously~\cite{Zhang2016Preserving}, by numerically calculating the effective magnetic field $\bm{\beta}_e$. The field $\bm{\beta}_e$ generates the same rotation operator as the CDD one, i.e.,
\begin{eqnarray}
e^{-i16\tau \bm{\beta}_e\cdot \bm{J}} = T: \prod_{j=1}^{16} U_j.
\end{eqnarray}

The maximal magnitude of the numerically obtained effective stray magnetic fields $\bm{\beta_e}$ is presented in the inset of Fig.~\ref{fig:dd}(d). As shown in the figure, the maximum of the effective stray magnetic fields $\beta_{e,m}=\max(|\bm{\beta_e}|)$ decreases as the bias $h$ increases, except $h/\beta_{e,m}<1$ where $\beta_{e,m}$ equals its largest possible value $\pi/(16\tau)$. With these effective stray magnetic fields at hand, we further calculate the spin squeezing parameter by employing Eq.~(\ref{eq:xi}) and also present the results in Fig.~\ref{fig:dd}(d). Note that the analytical Eq.~(\ref{eq:xi}) is invalid if $\beta_{e,m} < \beta_H$ or $t>t_s$ where $t_s$ denotes the optimal spin squeezing moment, because the assumption of a constant $J_z$ is violated. Clearly, the analytical results agree quite well with numerical ones. Such a good agreement justifies the validity of the approximations we make.

\section*{Conclusion and outlook}
\label{sec:con}

With numerical and analytical methods, we explore the spin squeezing dynamics of a ferromagnetic dipolar spin-1 BEC in stray magnetic fields. We find the spin squeezing dynamics of the condensate through the driven-OAT is severely destroyed by the stray magnetic fields. However, by introducing the CDD pulse sequence and a bias magnetic field, the spin squeezing dynamics is recovered almost perfectly. Our results point out a practical way to realize the optimal spin squeezing in a dipolar spin-1 BEC and can be employed to find a way to reach the Heisenberg limit quantum metrology.

In our calculation, although the stray magnetic fields are randomly distributed in both strength and direction, they are static. To be more practical, one may consider further time-dependent stray fields, pulse duration, and pulse imperfection, which are worth exploring in the future.

\bibliography{dss}

\section*{Acknowledgments}
W. Z. thanks M. S. Chapman for inspiring discussions on the spin squeezing. We thank the Beijing Computational Science Research Center for the hospitality at the early stage of this project. This work is supported by the National Natural Science Foundation of China Grant Nos. 11574239, 11434011, and 11275139, the National Basic Research Program of China Grant No. 2013CB922003.

\section*{Author contributions statement}

W.Z., P.X. and S.Y. conceived the idea. P. X. performed the calculations. P.X., H.S. and W.Z. analysed the results. P.X. and W.Z. wrote the manuscript. All authors reviewed the manuscript.

\section*{Additional information}

\textbf{Competing financial interests}

The author(s) declare no competing financial interests.

\end{document}